\documentclass[aps,pra,twocolumn,groupedaddress,showpacs]{revtex4}
\usepackage{graphicx}
\usepackage{dcolumn}
\usepackage{bm}
\begin{document}

\title{
Effectiveness of entangled coherent state in quantum metrology
}

\author{Osamu Hirota}
\email{hirota@lab.tamagawa.ac.jp}
\affiliation{
 Quantum ICT Research Institute, Tamagawa University, \\
6-1-1, Tamagawa Gakuen, Machida 194-8610 Tokyo, JAPAN
}%
\author{Dan Murakami}
 \email{mrkmd3ec@engs.tamagawa.ac.jp}
\affiliation{
Graduate School of  Tamagawa University, \\
6-1-1, Tamagawa Gakuen, Machida 194-8610 Tokyo, JAPAN
}%
\author{Kentaro Kato}
\email{kkatop@lab.tamagawa.ac.jp}
\affiliation{
 Quantum ICT Research Institute, Tamagawa University, \\
6-1-1, Tamagawa Gakuen, Machida 194-8610 Tokyo, JAPAN
}%

\date{\today}

\begin{abstract}
This Letter verifies the potential of several classes of 
entangled coherent state in well known quantum metrology which includes detection of 
classical external force, and shows that there is a class of entangled coherent state for the external force detection system  without the quantum limit in the detection of the light. 
In the case of the precision measurement of continuos parameter like phase measurement, the entangled coherent state with perfect entanglement does not provide remarkable benefit, 
but we provide a concrete example that certain class of entangled coherent state gives a remarkable sensitive detection scheme in the discrimination of digital signal affected by external force.
\end{abstract}

                             
\maketitle

\section{Introduction}
A generic interferometer has a shot noise limited sensitivity that scales with 
$\frac{1}{\sqrt{N}}$[1]. Here $N$ is the average photon number of the light source. 
This is called standard quantum limit.  Quantum communication theory including 
quantum estimation and quantum detection  may predict a possibility to beat such a limit. 
In fact, the above example of the limit corresponds to a subject of a phase estimation 
through quantum measurement. One can discuss the precision limits of quantum phase measurements by the Quantum Cramer-Rao inequality[2,3].
Since the mathematical treatment of the lower bound in physical problems has been 
clarified[4], the best resource for the phase estimation has been discussed[5,6]. 
Consequently when we prepare quantum correlations between the photons such as 
entangled state of light, the interferometer sensitivity can be improved 
by a factor of $\sqrt{N}$. That is, the sensitivity now scales with $\frac{1}{N}$ 
by employing the NOON state[5,6]. 
This limit is imposed by the Heisenberg uncertainty principle. 
The above Heisenberg limit is believed to be the ultimate precision in optical phase
estimation. Recently, more attractive feature has been discovered by employing entangled coherent state for the phase estimation problem by J.Joo et al [7], following 
the pioneering work by C.Gerry et al [8]. 
This work may open a new way. In fact, they showed that a specific type of 
entangled coherent state gives the smallest variance in the phase parameter 
in comparison to NOON state. The state is 
\begin{equation}
|\Psi\rangle_J = h_J (|\alpha_J \rangle_A|0 \rangle_B+|0 \rangle_A|\alpha_J \rangle_B )
\end{equation}
where $h_J$ is the normalization coefficient. This is a special case of the general entangled coherent state [9].
On the other hand, applications of the entangled state to discrete signals such as target 
detection or quantum reading have been discussed[10,11]. Thus, it is important that one examines a potential of a general ECS which is third way of entanglement state application.

The quasi Bell state based on entangled coherent states have a potential 
of the perfect entanglement[12], and applications to quantum teleportation and computation have been discussed[13,14,15,16]. Furthermore the feasibility [17] and experimental demonstration [18] of such states have been reported.

Thus, we are concerned with the effectiveness of the entangled coherent state in quantum metrology where various new technologies can emerge.
In this Letter, we consider the performance of several classes of 
entangled coherent state in well known quantum metrology which includes detection of 
classical external force, and propose a new method to detect the weak external force.

\section{Theoretical evaluation of physical limitation}
We can separate the issue of the theoretical limit on the ultimate sensitive measurement
into signals of continuous parameter and discrete parameter. The former is to clarify the precision of the measurement observable like the phase, and the later is to clarify the minimum error performance of the decision for signals in the radar or weak external force.
The quantum Cramer-Rao inequality has been formulated[2,3], in which the  bound is asymptotically achieved by the maximum likelihood estimator as well as the classical estimation theory. 

Here let $\rho (\theta)$ be the density operator of the system.
The estimation bound is given as follows:
First, the following operator equation is defined
\begin{equation}
\frac{\partial{\rho (\theta)}}{\partial{\theta}}=
\frac{1}{2}[A\rho (\theta)+\rho (\theta)A]
\end{equation}
where $A$ is called symmetric logarithmic derivative, which is self-adjoint operator.
Then the bound is given by
\begin{equation}
(\delta\theta)^2 \ge \frac{1}{F_Q}=\frac{1}{Tr(\rho (\theta)A^2)}
\end{equation}
$F_Q$ is also called quantum Fisher information.
The above inequality defines the principally smallest possible uncertainty in estimation 
of the value of phase.

On the other hand, in the radar detection or weak force detection, the problem becomes whether 
signal exists or not.
The formulation of the ultimate detection performance is called quantum
detection theory[2,3].  The limitation can be evaluated as follows:
\begin{eqnarray}
P_e &=& \xi_0 Tr\rho_0\Pi_1 + \xi_1 Tr\rho_1\Pi_0 \nonumber \\
& &\Pi_0+\Pi_1=I,\quad \Pi_i \ge 0
\end{eqnarray}
where $\{\xi_i \}$  is a priori probability of quantum states $\{\rho_i\}$ of the system, 
$\{\Pi_i\}$ is the detection operator, respectively.
 In the case of pure states, the optimum solution is
\begin{equation}
P_e=\frac{1}{2}[1-\sqrt {1-4\xi_0\xi_1|<\psi_0|\psi_1>|^2}]
\end{equation}

In the issue of precision measurement or super sensitive discrimination like digital acoustic laser microphone, the origins of the signals in our model are described as follows:\\
(a) Phase shift:$U(\theta|g)=exp(-\theta {a_A}^{\dagger}a_A)$\\
(b) Amplitude shift: $D(\alpha_s|g)=exp(\alpha_s {a_A}^{\dagger} - \alpha_s a_A)$,\\
where $g$ is an external force, $a_A$ and ${a_A}^{\dagger}$ are the annihilation and creation  operator of the the mode A, when an entangled state is applied.

\section{Short survey of entangled coherent state}
The four entangled state  so called quasi Bell state based on entangled coherent state are defined as follows:
\begin{eqnarray}
\left\{
\begin{array}{lcl}
|\Psi_1 \rangle &=& h_{1} (|\alpha \rangle_A|\alpha \rangle_B
+|-\alpha \rangle_A|-\alpha \rangle_B ) \\

|\Psi_2 \rangle &=& h_{2} (|\alpha \rangle_A|\alpha \rangle_B
-|-\alpha \rangle_A|-\alpha \rangle_B )\\

|\Psi_3 \rangle &=& h_{3} (|\alpha \rangle_A|-\alpha \rangle_B
+ |-\alpha \rangle_A|\alpha \rangle_B )\\

|\Psi_4 \rangle &=& h_{4} (|\alpha \rangle_A|-\alpha \rangle_B
-|-\alpha \rangle_A|\alpha \rangle_B)

\end{array}
\right.
\end{eqnarray}
where
$\{h_{i}\}$ are normalized constant:$h_{1}=h_{3}=1/\sqrt{2(1+\kappa^{2})}$,
$h_{2}=h_{4}=1/\sqrt{2(1-\kappa^{2})}$, and where
$ \langle \alpha | -\alpha \rangle = \kappa$ and
$\langle -\alpha | \alpha \rangle = \kappa^*$.

Some of these quasi Bell states are not orthogonal each other.
Here, if $\kappa = \kappa^*$, then the Gram matrix of them
becomes very simple as follows:
\begin{equation}
G=
\left(
\begin{array}{cccc}
1& 0& D& 0\\
0& 1& 0& 0\\
D& 0& 1& 0\\
0& 0& 0& 1\\
\end{array}
\right)
\label{ohmsgrammat}
\end{equation}
where $D=\frac{2 \kappa}{1 + {\kappa}^2}$.

  The degrees of entanglement for quasi Bell state are well known as follows:
 \begin{eqnarray}
& &E(|\Psi_1\rangle)=E(|\Psi_3\rangle) \\
                 &=& - \frac{1+C_{13}}{2} \log \frac{1+C_{13}}{2}
                   - \frac{1-C_{13}}{2} \log \frac{1-C_{13}}{2}\nonumber
\end{eqnarray}
where $E()$ is the entanglement of formation,  $C_{ij}=|\langle\Psi_i|\Psi_j\rangle|$, 
and $E(|\Psi_2\rangle)=E(|\Psi_4\rangle)=1$. 
Thus $|\Psi_2\rangle, |\Psi_4\rangle$ have the perfect entanglement [12].

\section{Quantitative properties of limitation}
\subsection{Phase estimation}
Recently, many authors applied the quantum Cramer-Rao bound to clarify the ultimate estimation of phase or phase shift of light. 
When an entangled  state  is prepared, considering the situation with no loss, a phase estimation bound is analytically given as follows[6,7]:
\begin{equation}
F_Q=4 [Tr_A \rho ({a_A}^{\dagger}a_A)^2 - (Tr_A\rho ({a_A}^{\dagger}a_A))^2 ]
\end{equation}
One can see from the above that NOON state gives the Heisenberg limit.
When one of four entangled coherent states is employed as light source, one can directly calculate the variance and the quantum Fisher information. As a result, the bound is given by

\begin{equation}
\delta\theta=\frac{1}{2\sqrt{[|\alpha|^4 K(1-K)+K|\alpha|^2]}}
\end{equation}
where
\begin{equation}
K_{\pm}=\frac{1\pm exp{(-2|\alpha|^2)}}{1\pm exp{(-4|\alpha|^2)}}
\end{equation}
where $+$ is for $|\Psi_1 \rangle$, and $|\Psi_3 \rangle$, and $-$ is for 
$|\Psi_2 \rangle$ and $|\Psi_4 \rangle$.

These bounds are larger than that of the state of Eq(1), 
regardless of the large entanglement. In fact, the estimation bound of Eq(1) was given 
as follows [7]:
\begin{equation}
\delta\theta=\frac{1}{2|\alpha_J|h_J\sqrt{(|\alpha_J|^2 +1) -h^2|\alpha_J|^2}}
\end{equation}
where $h_J=1/\sqrt {2(1+exp(-|\alpha_J|^2))}$,
which provides similar performance with NOON state and is superior to it in the case of 
weak amplitude $\alpha_J$ under the same total energy constraint. 
In addition, the energy of the mode A of $|\Psi\rangle_J$ is less than 
that of the quasi Bell state. 

On the other hand, by using superposed coherent state, it was shown that it allows displacement measurement at the Heisenberg limit[19]. 

A reason of the above benefit may come from the fact that the entangled coherent state of type of Eq(1) is interpreted as a superposition of NOON states, while the conventional entangled coherent states have more complicated structure. The precise physical meaning on such a difference will be discussed in the subsequent article.

\subsection{Detection of discrete external force}
There is an anther possibility of the precise measurement of the external force by quantum scheme. It is the discrete external force detection. In this case, we are concerned with whether certain entangled state can overcome the limit of the case of continuous precise measurement or not.
The problem goes to the detection of the event.
Here event means the fact in physical phenomena like yes or no. There are many such detection problems in communication theory, for example, the coherent laser radar or the detection of parameter shift of a light by an external force in interferometer.

We here restrict the problem to the binary detection. So detection targets are two quantum states. We take an energy shift by external force in this Letter. That is, the problem is to decide the shift from the steady state by the external force.
Our proposal to realize good performance is as follows:

Let us assume that  following entangled coherent state is employed, 
\begin{equation}
|\Psi (0) \rangle = h(0) (|\alpha \rangle_A|\beta \rangle_B -|-\alpha \rangle_A|0 \rangle_B ) 
\end{equation}
We design the system such that the energy of the mode B is reduced when the external force is applied to the system. The energy shift is denoted as $\epsilon$. 
However, the external force does not affect to the vacuum state.
As a result, after the external force is applied to the system, 
the state of the system is as follows:
\begin{equation}
|\Psi (1)\rangle = h(1) (|\alpha \rangle_A|\beta - \sqrt \epsilon \rangle_B 
-|-\alpha \rangle_A|0 \rangle_B ) 
\end{equation}
Here we design that $\beta - \sqrt \epsilon=0$. So we have the following situation.
The inner product between two entangled coherent states becomes 
\begin{eqnarray}
& & \langle\Psi(1)|\Psi(0)\rangle=
 h_0 h(1) (\langle \alpha|_A \langle \beta |_B -\langle -\alpha|_A \langle 0 |_B )
\nonumber \\
& & \times (|\alpha \rangle_A|0 \rangle_B -|-\alpha \rangle_A|0 \rangle_B )=0
\end{eqnarray}
Thus, the inner product is independent of 
the amplitude parameter of the coherent state, and the performance is given by Eq(5) and the above. Thus the detection performance is independent of the energy of the light source, but 
$\epsilon$ depends on the system design for external force detection.

This type of detection is sometimes called threshold detection problem in the communication theory. 

Secondly, we examine the case of the single mode coherent state.
The initial state and affected state are as follows:
\begin{eqnarray}
|\alpha(0) \rangle &=& |\beta \rangle \\
|\alpha(1)\rangle &=& |\beta- \sqrt \epsilon \rangle
\end{eqnarray}
When $\beta - \sqrt \epsilon$ is 0, 
The inner product of two coherent states is
\begin{equation}
\langle \beta|0 \rangle =exp{(-|\beta|^2/2)}
\end{equation}

According to Eq(5), we have 
\begin{eqnarray}
P_e(C)&=&\frac{1}{2}[1-\sqrt {1-4\xi_0\xi_1 exp(-|\beta|^2)}]\nonumber \\
P_e(ECS)&=& 0
\end{eqnarray}

In several applications, we encounter a situation that a priori probabilities of the external force as signals are unknown.
Then the detection scheme becomes quantum minimax detection[20]. However, the performance is given by putting $\xi_0=\xi_1=1/2$ in the above equations, because these are the worst a priori probabilities in any case of the binary pure state situation.

Furthermore, in the external force detection, we should consider the case of unknown shift by the external force. We suggest the $M$ parallel systems of the above system which are designed by following energy attenuation:
\begin{eqnarray}
&&\beta_1 - \sqrt \epsilon_1=0\nonumber \\
&&\beta_2 - \sqrt \epsilon_2=0\nonumber \\
&&\vdots \nonumber \\
&&\beta_M - \sqrt \epsilon_M=0
\end{eqnarray}
When one of systems clicks the result of the external force, one can record 
the external force  detection.

\section{CONCLUSION}
We have discussed some applications of several  classes of entangled coherent state to quantum metrology. The role of entangled coherent state with the perfect entanglement 
is minor in the precision measurement of the continuous parameter.  
However, certain type of entangled coherent state provides a great advantage in the discrimination of the digital parameter affected by external force.
Following the above results, we have given a new method to detect the weak external force
by using the parallel system of digital detection.

\begin{acknowledgments}
We are grateful Bill Munro and T.S.Usuda for helpful discussions, and acknowledge the president Y.Obara of Tamagawa University for special funding.
\end{acknowledgments}

\end{document}